# THE ECONOMIC COSTS OF THE RUSSIA-UKRAINE WAR: A SYNTHETIC CONTROL STUDY OF (LOST) ENTREPRENEURSHIP

D. Audretsch[1], P. P. Momtaz[2,3,4]*, H. Motuzenko[5], S. Vismara[6]

**Affiliations:**

[1]O'Neill School of Public & Environmental Affairs, Indiana University, Bloomington 47405, USA

[2]University of California, Los Angeles (UCLA), Anderson School of Management, 110 Westwood Plaza, CA 90095, Los Angeles, United States.

[3]TUM School of Management, Arcisstr. 21, 80333 München, Germany.

[4]University College London, Computer Science, Centre for Blockchain Technologies, Malet Place, London, United Kingdom.

[5]Goethe University Frankfurt, Department of Business and Economics, Theodor-W-Adorno-Platz 3, 60323 Frankfurt, Germany.

[6]University of Bergamo, Department of Management, Via dei Caniana, 2, 24127 Bergamo, Italy.

*Correspondence to: momtaz@ucla.edu.

# THE ECONOMIC COSTS OF THE RUSSIA-UKRAINE WAR:
# A SYNTHETIC CONTROL STUDY OF (LOST) ENTREPRENEURSHIP

**Abstract:**

This synthetic control study quantifies the economic costs of the Russo-Ukrainian war in terms of foregone entrepreneurial activity in both countries since the invasion of Crimea in 2014. Relative to its synthetic counterfactual, Ukraine's number of self-employed dropped by 675,000, corresponding to a relative loss of 20%. The number of Ukrainian SMEs temporarily dropped by 71,000 (14%) and recovered within five years of the conflict. In contrast, Russia had lost more than 1.4 million SMEs (42%) five years into the conflict. The disappearance of Russian SMEs is driven by both fewer new businesses created and more existing business closures.

1.      **Introduction**

Is war good for the economy? A deeply ingrained conventional wisdom suggests the affirmative: "*One of the more enduring myths in Western society is that wars are somehow good for the economy*" (Thought Co., 2018) and, similarly, "*[o]ne of the enduring beliefs of modern times is that war and its associated military spending has created positive economic outcomes*" (Institute for Economics & Peace, 2011). In contrast, empirical evidence of the economic costs of conflict in terms of the GDP per capita suggests a negative effect. The cross-country study by Costalli et al. (2017) estimates that war reduces GDP on average by 17.5%. Existing studies attribute the negative effect of war on GDP to trade disruptions, private investment suspensions, human capital losses, physical capital destruction, technological regress, political instability, and general uncertainty (Abadie and Gardeazabal, 2003; Alesina and Perotti, 1996; Barro, 1991; Glick and Taylor, 2010). This literature has evoked criticism that the focus on GDP as a state variable is not fully informative about conflicts' longer-term consequences for economic dynamics, such as entrepreneurial activity. For example, in their seminal study of the economic costs of conflict, Abadie and Gardeazabal (2003, p. 113) discuss that "entrepreneurs […] had been specific targets of violence and extortion […] However, little research has been carried out to assess the economic effects". Very few studies have indeed so far dealt explicitly with the impact of conflict on entrepreneurship.

Although a few studies have now started to explore how conflict might impact entrepreneurial activity from various perspectives, one overarching question has not yet been addressed: Does violent conflict between two countries impact *aggregate* entrepreneurship in those countries? Existing studies often investigate civil strife and hence focus on specific regimes or sectors in conflict-ridden countries rather than on the conflict's aggregate impact, or they focus on specific phenomena, such as necessity or social entrepreneurship (Bozzoli et al., 2013; Brück et al., 2011; Tobias et al., 2016; Collier, 1999; Gimenez-Nadal et al., 2019; Bauer et al., 2016; Voors et al., 2012). The resulting research gap is striking given that policymakers and industry associations have long warned about potentially devastating consequences of violent conflict on economic growth through foregone entrepreneurship (Astrov et al., 2022; European Business Association, 2022).

Consistent with conventional wisdom and contrary to macroeconomic evidence, the majority of the studies report a *positive* relation between violent conflict and entrepreneurial activity (e.g., Bozzoli et al., 2013; Ciarli et al., 2015; Brück et al., 2011), with only a few rebutting studies (e.g., Camacho and Rodriguez, 2012; Deininger, 2003). However, there are concerns about the identification of the conflicts' estimated negative effects on entrepreneurial activity. First, most studies are qualitative and develop a narrative rather than an empirical test of the war-entrepreneurship relation (Brück et al., 2011). They develop local "unit theories" instead of global "programmatic theories" that are of little use "to make clearer and more useful recommendations to leaders and policy-makers" (Aguinis et al., 2022, p. 1671). Second, those studies that are quantitative, are often weakly or not at all identified, with only a few exceptions (e.g., Bozzoli, 2013, Ciarli et al., 2015; Camacho and Rodriguez, 2012). A common problem with many empirical studies is that they examine enduring conflicts over truncated sample periods. For example, several studies focus on the domestic strife in Colombia, which started in the 1960s, but consider more recent sample periods that are shorter by an arbitrary amount (e.g., Bozzoli et al., 2013; Gimenez et al., 2019). Any such identified effect might not be causal, but rather reflect a "recovery effect" that is documented in post-conflict studies (e.g., Tobias et al., 2016), which could be a root cause for the many (possibly misidentified) positive effects of war in the literature.

To overcome this challenge, we explore the Russo-Ukrainian war that erupted with the Russian occupation of the Ukrainian peninsula of Crimea in February 2014 as a natural experiment. The war suggests itself as a nearly ideal pre-versus-post comparison because the start of the conflict is widely recognized as an exogenous shock, and not endogenously driven by socio-economic tensions between the parties in conflict like many other violent conflicts. Russia and Ukraine have been major trading partners ever since the dissolution of the Soviet Union. As Korovkin and Makarin (2023, p. 8) state, "[t]he decision to occupy Crimea was made secretly by Vladimir Putin and a handful of senior security advisors; it took everyone else by surprise." To ensure casual identification of our pre- versus post-war comparison of entrepreneurial activity in Ukraine and Russia, we construct counterfactuals with the Synthetic Control Method (SCM) (Abadie et al., 2010; 2015; Chen et al., 2023). The SCM has been described as "the most important innovation in the policy evaluation literature in the last 15 years" (Athey and Imbens, 2017, p. 9) and is the only viable quasi-experimental method in our empirical context because entrepreneurial activity in Ukraine, Russia, and the control countries have heterogeneous levels and follow non-parallel pre-trends. Thus, only synthetic versions of Ukraine and Russia that mimic entrepreneurial trajectories in these countries are admissible counterfactuals to estimate the causal effect of war.

We examine entrepreneurial activity over a symmetric eleven-year event window around the start of the conflict in 2014. Our balanced annual country-level panel includes Ukraine, Russia, and 45 donor countries. We consider entrepreneurial activity in these countries along two distinct measures. First, we consider self-employment from nationally representative labor force surveys as the percentage of self-employed of the employed population. Second, we consider SMEs from national registry data as the percentage of registered LLCs of the working-age population. These are standard proxies used to measure entrepreneurship and are complementary. While the percentage of registered SMEs only reflects the formal sector, the fraction of self-employed also captures the informal sector. The latter is important because the shadow economy accounts for more than a quarter of entrepreneurial activity in both Ukraine and Russia (Williams, 2009; Bauer et al., 2016).

Our empirical results suggest that the Russo-Ukrainian war had a negative effect on entrepreneurship in both countries. The effect varies in intensity and persistence. The percentage of self-employed in Ukraine dropped sharply by 19.5% in the first conflict year, corresponding to a war-induced unemployment of 675 thousand formerly self-employed Ukrainians. Although this effect partially subsides over time, foregone entrepreneurship proxied by the percentage of self-employed still accounts to 8% (279 thousand Ukrainians) in the fifth conflict year. We check the internal validity of our synthetic control through pre- versus post-war ratios of the mean squared prediction error between countries and in-space placebo tests, and we fail to falsify our model. Analogous to statistical significance tests, we calculate that the probability of estimating an effect as large as ours "by chance" is 3.3%. For self-employment in Russia, the SCM fails to construct a reasonable counterfactual.

The negative war effect is also observable in the fraction of SMEs, although Ukraine and Russia are affected differently. For Ukraine, the difference in SMEs with its synthetic version peaks in 2016, with a gap of 71 thousand SMEs that corresponds to a relative loss of 14%, and recovers back to the synthetic trajectory five years into the war. Implied p-values are only significant until 2016 (p-value 8.7%) and non-significant thereafter, reconfirming the temporary effect. For Russia, in contrast, the effect is persistent and reinforcing. Five years into the conflict, the number of Russian SMEs has dropped by 42.2%, accounting for an absolute loss of 1.4 million SMEs. Investigating the sources of the staggering disappearance of Russian SMEs, we find that it is caused by both fewer new business creations and, to a larger extent, more existing business closures.

## 2. Empirical Design and Results

### 2.1 Institutional background: The Russo-Ukrainian conflict

Ukraine is located in Eastern Europe, has a population of roughly 40 million inhabitants, and, with a geographic area of more than 600,000 square kilometers, it is the largest country contained within the borders of the European continent. Ukraine first gained sovereignty as an independent nation in 1917 became part of the Soviet Union after WWI. At the end of WWII, Soviet dictator Stalin negotiated with the United Nations to include Ukraine as Ukrainian Soviet Socialist Republic in the Soviet Union. Crimea was part of the Russian Soviet Federative Socialist Republic until 1954, when Nikita Khrushchev, First Secretary of the Communist Party of the Soviet Union, transferred it to the Ukrainian Soviet Socialist Republic at the 300th anniversary of the Treaty of Pereiaslav, an agreement that secured Cossack Hetmanate, a predecessor state of Ukraine, the military protection of the Tsardom of Russia. With the dissolution of the Soviet Union, Ukraine regained national sovereignty and reorganized Crimea as a de-jure autonomous republic in 1995. In February 2014, Russia invaded Crimea and organized a referendum declaring independence from Ukraine. As per UN Resolution 68/262 of March 27, 2014, most UN member states view the referendum as illegal and condemn the Russian occupation as a violation of the law of nations.

Russia-commanded armed para-military troops also subdued the municipal administrations of Ukrainian cities Donetsk and Luhansk in April 2014, initiating the war in Donbas (the farthest Eastern Ukrainian territory bordering Russia). The Office of the United Nations High Commissioner for Human Rights (OHCHR) declared the "total collapse of law and order" in the Donbas region in July 2014. Until Russia launched its full-scale armed invasion of Ukraine in February 2022, which characterizes the current state of the Russo-Ukrainian conflict, the United Nations (2022) estimate more than 14,000 casualties in the Donbas region, with more than 3,000 civilian victims. The number of displaced Ukrainians amounts to several millions. Importantly, Ukraine and Russia were major trading partners until the war, with tariffs set to zero. Ukrainian exports to and imports from Russia, which accounted for a quarter of all Ukrainian exports and a third of all Ukrainian imports, decreased from two years before to two years after the conflict by 61% and 60%, respectively.

## 2.2 Sample and Summary Statistics

To estimate the conflict's impact on entrepreneurship, we construct a balanced annual country-level panel for the 2009-2019 period for Ukraine, Russia, as well as for 45 "donor pool" countries, for which data on the two outcome variables and eight predictors of entrepreneurial activity are available. The donor pool includes all countries that are plausibly related to Ukraine and Russia either through the joint former membership in the Soviet Union or through their satellite status (e.g., Poland, Hungary, and Bulgaria), as well as other countries on the Eurasian continent and elsewhere. Table 2 below lists all donor countries that are used in the construction of synthetic versions of Ukraine and Russia.

We focus on two outcome variables to capture country-level entrepreneurial activity. First, we examine *the percentage of self-employed of total employment* from the *International Labor Organization's ILOSTAT database*[1], which gathers annual data on the number of self-employed from nationally representative labor force surveys. Second, we explore *the percentage of registered LLCs of the working-age population* from the *World Bank Entrepreneurship Database*[2], which gathers annual SME data from business registries. The self-employment variable measures entrepreneurship both in the *formal* and in the *informal* sector, while the SME variable measures entrepreneurship only in the formal sector.

---

[1] See, for further details, https://ilostat.ilo.org/topics/employment/, and retrieved from the World Bank (https://data.worldbank.org/indicator/SL.EMP.SELF.ZS) in June 2022.
[2] The data was retrieved from https://www.worldbank.org/en/programs/entrepreneurship#new, which collects data from national business registries, in June 2022. We manually checked the official statistics in Ukraine and Russia. We identified differences in the Ukrainian statistics for the number of LLCs in some years. In these cases, we decided to go with the Official Accounts of Ukraine.

Summary statistics for the two outcome variables for Ukraine, Russia, and the donor pool are in Table 1. The percentage of self-employed decreased from 18.64% to 15.80% from the pre- to the post-invasion period in Ukraine, with the 2.84% difference in means corresponding to 1.13 million Ukrainians being statistically highly significant. The average percentage of self-employed also decreased in Russia by 0.41% (0.18 million Russians) and in the donor pool by 1.46%, albeit statistically non-significant. In contrast, the percentage of registered SMEs increased statistically significantly by 0.37% (99.29 thousand Ukrainian SMEs), 0.36% (251.16 thousand Russian SMEs), and 1.21% (125.98 thousand SMEs in donor pool countries) in the post-invasion period.

**[PLEASE INSERT TABLE 1 ABOUT HERE]**

We explain these outcome variables with eight covariates, including GDP per capita, GDP growth, FDI net inflows, a proxy for the ease of starting a business, domestic credit to the private sector as a percentage of GDP, unemployment, R&D expenditure, and patent applications by residents. Variable definitions and summary statistics are Table A.1 in the Appendix.[3] The average percentages of self-employment and registered SMEs for the full sample (Russia, Ukraine, and the donor pool) are 5.68% (pre-invasion: 5.12%, post-invasion: 6.30%) and 22.45% (pre-invasion: 23.12%, post-invasion: 21.65%), which starkly contrast with those for Ukraine and Russia in Table 1, and therefore highlight the need to carefully select controls for the counterfactuals.

2.3    *Research Design: Synthetic Control Method*

The summary statistics in Table 1 might (misleadingly) suggest that the Russian invasion of Ukraine had a negative effect only on self-employment in Ukraine and positive effects on the number of registered SMEs in Ukraine and Russia. Such an inference might in reality just reflect ambiguity in how control countries were chosen and their ability to reproduce the outcome trajectory of the counterfactual, which is a common problem in comparative case studies (Abadie et al., 2010). Figure 1 illustrates the problem. Panel a shows the trends in the percentage of self-employed, and Panel b shows the trends in the percentage of registered SMEs over the 2009-2019 sample period. These plots suggest that the full donor pool is a poor control group for Ukraine and Russia because the levels in both variables vary significantly across the three groups and, even pre-invasion, the donor pool trends are not parallel to those of Ukraine and Russia. For example, for self-employment, the donor pool monotonically decreases at almost a constant rate from 25% to about 21% over the sample period, while Ukraine's trend is actually increasing in the pre-invasion period from 17% to 20%.

---
[3] We considered a battery of 18 covariates during the synthetic control construction process, but found the aforementioned eight covariates to lead to the most robust synthetic countries. The additional variables included inflation, mobile cellular subscriptions, branches of commercial banks, the resolving insolvency score, as well as the six World Bank Governance Indicators. All data was retrieved from World Bank in June 2022.

**[PLEASE INSERT FIGURE 1 ABOUT HERE]**

We address these problems with the SCM (Abadie and Gardeazabal, 2003; Abadie et al., 2010; Abadie et al., 2015). We construct synthetic Ukraine and Russia as linear convex combinations of countries in the donor pool that most closely resemble the countries in terms of the pre-invasion predictors of self-employment and number of registered SMEs. Self-employment for synthetic Ukraine is predicted by pre-invasion country-level characteristic means for (i) GDP growth, (ii) unemployment, and (iii) domestic credit. The synthetic version (i=-0.37%, ii=10.71%, iii=71.35) tracks Ukraine (i=-0.52%, ii=7.64%, iii=75.63) much better than the donor pool (i=1.66%, ii=9.48%, iii=85.1). The tracking error, measured as the absolute percentage deviation of Ukraine's self-employment predictor means between the synthetic version and the donor pool, reduces by almost 85% in synthetic Ukraine. In contrast, we are not able to construct a synthetic Russia that tracks the pre-invasion evolution trajectory of self-employment sufficiently well (see *Section 2.4.2*).

Pre-invasion percentages of registered SMEs in Ukraine and Russia can be tracked closely by (i) FDI, (ii) the World Bank's starting-a-business score, (iii) R&D expenditures, and (iv) residential patent applications. For Ukraine (i=$6.97 billion, ii=73.33, iii=0.80, iv=2,635), the synthetic version (i=$7.04 billion, ii=73.33, iii=0.80, iv=2,642) is much better than the donor pool average (i=$16.02 billion, ii=85.09, iii=1.44, iv=12,424). For Russia (i=$54.90 billion, ii=88.60, iii=1.08, iv=27,666), the synthetic version is comparable to the donor pool average. The synthetic countries are constructed based on the weights on donor pool countries shown in Table 2. Self-employment trends in pre-invasion Ukraine are best reproduced by Slovenia (42%), Estonia (37%), and Armenia (19%); SMEs trends in pre-invasion Ukraine are best reproduced by Tajikistan (45%), India (15%), Austria (11%), and SME trends in pre-invasion Russia are best reproduced by Denmark (36%), Serbia (22%), Bulgaria (20%), in addition to a number of donor countries with smaller weights.

**[PLEASE INSERT TABLE 2 ABOUT HERE]**

*2.4    Results*

*2.4.1    Self-employment: Ukraine versus Synthetic Ukraine*

Panel a of Figure 2 displays self-employment as a percentage of total employment for Ukraine and its synthetic counterpart during the 2009-2019 sample period. The pre-invasion trends for Ukraine and its synthetic version are reassuring. In contrast to the trend differences between Ukraine and the donor pool in Figure 1, self-employment in synthetic Ukraine closely tracks the trajectory of self-employment in Ukraine for the pre-invasion period. The negligible pre-invasion tracking error suggests that synthetic Ukraine represents a reasonable approximation to the percentage of Ukrainian self-employed had Russia not invaded Ukraine in 2014.

The estimated causal effect of the Russian invasion of Ukraine on self-employment is the difference between the percentage of self-employed in Ukraine and its synthetic equivalent in the post-invasion period. Immediately following the invasion, self-employment in Ukraine drops dramatically by 674.5 thousand from about 3.46 million to 2.61 million, corresponding to a relative loss of 19.5%, and remains at this level over the remaining sample period. Self-employment in synthetic Ukraine also decreases in the 2016-2019 period. The relative difference is decreasing, suggesting that self-employment in Ukraine recovered from the initial shock of the invasion in the post-invasion period, starting in the second post-invasion year in 2016. Panel b makes this notion explicit by plotting the annual gaps in self-employment between Ukraine and synthetic Ukraine, with the estimated effect for synthetic Ukraine normalized to zero. Panel b suggests that the absolute loss in self-employed Ukrainians gradually decreases in the following years to 492,995 (-14.3%) in year 2016; 362,896 (-10.5%) in 2017; 370,118 (-10.7%) in 2018; and 278,553 (-8.1%) in 2019. In the hypothetical scenario that the recovery path of self-employment in Ukraine would continue without interruption (i.e., ignoring the further escalation in 2022), Ukraine would break even with synthetic Ukraine nine years into the conflict.

To evaluate the significance of our estimated effect of the Russian invasion on self-employment in Ukraine, we cannot rely on statistical significance tests because the invasion lacks randomization (Abadie et al., 2010; 2015). Instead, a viable alternative approach to assess the significance of the identified effect is to explore whether the effect could be entirely driven by chance, using placebo tests. As in Abadie and Gardeazabal (2003), we estimate "in-space placebo tests" by applying the SCM to non-invaded non-invading countries in the donor pool. The placebo tests would falsify our interpretation that the Russian invasion caused the decrease in Ukrainian self-employment if they would lead to estimated gaps between synthetic and actual countries that are similar in magnitude to that estimated for Ukraine.

Panel c plots the placebo tests for donor-pool countries with Mean Squared Prediction Error (MSPE) between self-employment in an actual country and its synthetic version during the pre-invasion period of less than or equal to ten times that of Ukraine to ensure that the tests are not contaminated by synthetic versions of control countries that cannot be reproduced with our variables and/or donor pool (Abadie et al., 2010). The gray lines correspond to the gaps associated with individual donor pool countries, while the black line corresponds to the estimated annual self-employment gaps of Ukraine. Panel c suggests that the gap for Ukraine is relatively large compared to the gaps for the countries in the donor pool. The placebo tests indicate that the probability of obtaining an effect as large as the immediate effect of the Russian invasion on self-employment in Ukraine in 2015 when the invasion status is reassigned at random in our data is 1/30 = 0.033 (i.e., the number of placebo effects exceeding that of Ukraine in 2015 divided by the number of all estimated effects; see Abadie et al., 2015, p. 500). Finally, in unreported results, we find that the ratio of post- and pre-invasion MSPEs for Ukraine exceeds that of the donor-pool countries. The post-invasion MSPE is more than 100 times as large as that for the pre-invasion period for Ukraine. No country in the donor pool has a larger MSPE ratio, reconfirming the significance of the war's effect on self-employment in Ukraine.

**[PLEASE INSERT FIGURE 2 ABOUT HERE]**

*2.4.2   Self-employment: Russia versus Synthetic Russia*

We also tried to estimate the effect of the invasion on self-employment in Russia. However, no combination of our donor pool countries reproduces Russian self-employment in the pre-invasion period reasonably well. Therefore, for brevity, the analysis is shown in Figure IA.1 in the Internet Appendix.

*2.4.3   SMEs: Ukraine versus synthetic Ukraine*

Turning to the invasion's effect on registered SMEs, we first discuss the evidence for Ukraine. Figure 3 shows trends in the percentage of SMEs relative to the working-age population for Ukraine and its synthetic version, the annual gap between the two, and placebo tests. Again, there is little variation in the pre-invasion period. Compared to the trend differences between Ukraine and the donor pool in Figure 1, the evolution of SMEs in synthetic Ukraine closely tracks that of Ukraine in the pre-invasion period. Trends diverge in the year of the invasion. The annual gap reaches its maximum two years after the invasion and then climbs back to the counterfactual trajectory for the synthetic country in 2019, indicating a full recovery period of three years. The 2016 effect peak corresponds to a temporary dip in SMEs of 14%. In absolute terms, the total number of registered SMEs in Ukraine is 489,626 and the implied number for synthetic Ukraine is 560,378, which indicates that the Russian invasion cost Ukraine 70,752 registered SMEs by 2016.

In-space placebo effects for all countries in the donor pool with a pre- versus post-MSPE ratio below 10 (leaving 23 countries) are in Panel c. The probability of obtaining an effect larger than that estimated for Ukraine in 2016 by randomly assigning the invasion status to countries in the donor pool is 8.7% (=2/23). In contrast, the corresponding probability for the year 2019 is 56.5% (=13/23). The placebo tests thus support our inference that the Russian invasion had a temporary and significant effect on the number of registered SMEs in Ukraine in 2016, but the effect did not persist, was non-significant with a probability of 26.1% (=6/23) in 2017, and fully subsided over the next three years.

*2.4.4    SMEs: Russia versus synthetic Russia*

The patterns for registered SMEs as a percentage of the working-age population in Russia contrast with those estimated in Ukraine. Panels a and b of Figure 3 indicate that the Russian invasion of Ukraine did not have an immediate impact on SMEs in Russia but a very strong impact after the first two years following the invasion. Pre-invasion trends between Russia and its synthetic version a very similar. These trends slightly diverge in the two post-invasion years, but the difference is not significant. The annual gap in 2016 has a probability of being estimated at random of 21.6% according to the placebo tests in Panel c. The percentage of registered SMEs in Russia decreases dramatically over the 2017 to 2019 period. The placebo tests suggest that the chance of obtaining an effect as large as the 2019 one at random is 8.1% (=3/37). Therefore, these results suggest that the effect on Russian SMEs lags but is very pronounced in the long term, with potentially even higher rates after the end of our sample period in 2019. As for the absolute values, in the last year of our analysis, the number of Russian SMEs was 3.3 million, whereas according to synthetic Russia this number would have been 4.8 million. Thus, in the absence of war, the implied gap of 1.4 million corresponds to a relative loss of 42%.

**[PLEASE INSERT FIGURE 3 ABOUT HERE]**

*2.5    Additional Results: Birth and death rates of Russian SMEs*

The *World Bank Entrepreneurship Database* also provides information on birth and death rates of registered SMEs only for Russia. We leverage the data here to shed light on the mechanics behind the disappearance of SMEs in Russia. Unfortunately, the data are not available for Ukraine. For both birth and death rates of Russian SMEs, we create synthetic versions of Russia and compare the post-invasion trends in Figure 4. At the end of our sample period in 2019, the Russian SME birth rate is down one-third and the death rate is increased by a factor of 2.5 relative to the synthetic versions, indicating that the disappearance of Russian SMEs is driven by both fewer newly created businesses and (to a larger extent) more existing business closures.

**[PLEASE INSERT FIGURE 4 ABOUT HERE]**

## 3. Discussion and Concluding Remarks

Does violent conflict impact entrepreneurship? Although the question has started to attract attention, the *aggregate* impact of war on entrepreneurial activity in conflict-ridden countries is underexplored. Motivated by this research gap, we investigate how the Russo-Ukrainian war that erupted with the Russian occupation of the Ukrainian peninsula of Crimea in February 2014 affected aggregate counts of self-employed and registered SMEs in Ukraine and Russia. Employing the Synthetic Control Method (SCM) to construct synthetic versions of Ukraine and Russia as counterfactuals, we find that the war had a dramatic negative effect on self-employment as a percentage of total employment in Ukraine. The fraction of self-employed dropped by 19.5% when the conflict erupted, corresponding to 674.5 thousand Ukrainians. Although Ukrainian self-employment partially recovered subsequently, we estimate that foregone entrepreneurship associated with solo-entrepreneurs still amounted to 8% five years into the conflict. Our evidence for SMEs sheds additional light on the war's impact. The war hit Ukrainian SMEs only temporarily. Although 70.8 thousand SMEs accounting for 14% of all Ukrainian SMEs were lost immediately following the conflict's outbreak, Ukraine recovered fully within five years. In contrast, Russian SMEs were hit stronger and never recovered. Five years into the conflict, Russia had foregone entrepreneurial activity in the amount of 1.4 million SMEs, a staggering 42% loss. Further, investigating the sources of the disappearance of Russian SMEs, we show that both fewer new business creation and (to a larger extent) more existing business closures drive the effect.

To the best of our knowledge, this is the first synthetic control study that quantifies the economic costs of conflict in terms of aggregate foregone entrepreneurship. Our study contributes to three distinct streams in the literature. First, we contribute to the growing evidence on the economic costs of conflict (e.g., Abadie and Gardeazabal, 2003; de Groot et al., 2022) by expanding the set of examined outcomes from GDP per capita to lost entrepreneurship. Given that the contributions of entrepreneurship to GDP are heterogeneous across countries – e.g., entrepreneurship's value added is 20% in Russia and 53% in Ukraine (Statista, 2022; European Commission, 2021) – our results provide an explanation for why war impacts countries' GDPs so differently. Further, given that nascent entrepreneurship's value added materializes with a lag between starting up and being profitable, our study also helps explain why GDP recovers with delays after conflicts (Costalli et al., 2017). Second, our study contributes to the literature on conflict and entrepreneurship that has so far focused on specific aspects of how conflict impacts entrepreneurship, such as displacements and necessity entrepreneurship (e.g., Bozzoli et al., 2013) and social entrepreneurship (Gimenez-Nadal et al., 2019; Bauer et al, 2016). Our study provides an account of the aggregate negative impact of the Russia-Ukraine conflict on entrepreneurship in both countries and challenges the majority of evidence arguing that conflict impacts entrepreneurship in a positive way. Third, we contribute to the recent body of work on the Russo-Ukrainian war (Astrov et al., 2022, for a review). While no study has studied the entrepreneurial ramifications of the war so far, existing work quantifies the economic costs of the war and potential sources. In particular, our work is closely related to Bluszcz and Valente's (2022) synthetic control study of Ukrainian GDP, whose estimated effects are similar to ours, as well as Korovkin and Makarin (2023) and Hoffmann and Neuenkirch (2017) who show that sources of decreased GDP include trade and financial market frictions, respectively. Further, the disappearance of entrepreneurial activity might be one factor behind the identified mental health consequences of the Russo-Ukrainian war (Bai et al., 2022; Chudzicka-Czupała et al., 2023; Jawaid et al., 2022; Kossowska et al., 2023).

Our empirical analyses are not without limitations. First, the SCM identifying assumption is no interference between countries ("no spillover assumption"). However, it is possible that the Russo-Ukrainian war affected entrepreneurship also in other countries, e.g., by deterring international ventures (Zahra, 2022) and cross-border activity (Welter et al., 2018). Second, the absence of war casualties and displacements in donor pool countries might bias the synthetic controls. Note, however, that both caveats would cause our estimates to be conservative. Additionally, we failed to construct a synthetic version of Russia with respect to the self-employment variable (possibly due to the idiosyncratic system of "blat" in Russia, see Aidis et al., 2008; Puffer et al., 2010).

Finally, our results have potentially important implications for entrepreneurs and policymakers. The overall conclusion from our study is that the Russo-Ukrainian war inhibits entrepreneurial activity, which results from both lower birth and higher death rates of SMEs. First, existing entrepreneurs would need to improve their resilience to the increased risk of failure, inter alia, by stabilizing demand through diversifying into new markets and increasing their resilience to supply chain shocks. Policymakers can support existing entrepreneurs during the Russia-Ukraine conflict, e.g., by managing energy costs and ensuring a steady supply of input factors. Second, to mitigate the war's negative impact on the birth rate of new ventures, policymakers could nudge potential entrepreneurs into venturing, e.g., through subsidy programs. Understanding the various support programs available to policymakers to mitigate the detrimental consequences of war on entrepreneurship is an important open question that merits further research. This would also involve an investigation of the potential mechanisms behind the conflict's negative effect on entrepreneurship, such as the relative importance of operational, financial, and human capital-related constraints at times of war, which is left to future research. While there remains much for future research to substantiate, the results of this paper suggest that the loss of entrepreneurial activity in the Russia-Ukraine war may take generations to overcome.

**Exhibits**

**Table 1: Pre- and post-conflict entrepreneurship in Ukraine, Russia, and the donor pool**

|  | Pre-invasion | | Post-invasion | | Difference |
|---|---|---|---|---|---|
|  | Mean | St. Dev. | Mean | St. Dev. | in means |
| *Panel a: Ukraine* | | | | | |
| Self-employed, in million | 3.71 | 0.04 | 2.59 | 0.2 | -1.13*** |
| Self-employed, in % of total employment | 18.64 | 0.72 | 15.80 | 0.12 | -2.84*** |
| SMEs, in thousand | 450.58 | 51.15 | 549.87 | 55.85 | 99.29** |
| SMEs, in % of working-age population | 1.41 | 0.17 | 1.78 | 0.21 | 0.37** |
| *Panel b: Russia* | | | | | |
| Self-employed, in million | 5.28 | 0.05 | 5.10 | 0.25 | -0.18 |
| Self-employed, in % of total employment | 7.47 | 0.78 | 7.06 | 0.34 | -0.41 |
| SMEs, in thousand | 3,430.6 | 257.71 | 3,681.75 | 233.77 | 251.16 |
| SMEs, in % of working-age population | 3.35 | 0.26 | 3.71 | 0.19 | 0.36** |
| *Panel c: Donor pool* | | | | | |
| Self-employed, in million | 8.79 | 45.17 | 8.12 | 40.63 | -6.69 |
| Self-employed, in % of total employment | 23.57 | 17.25 | 22.10 | 16.17 | -1.46 |
| SMEs, in thousand | 439.12 | 613.45 | 565.1 | 787.05 | 125.98* |
| SMEs, in % of working-age population | 5.24 | 5.15 | 6.46 | 5.32 | 1.21** |

**Figure 1: Non-parallel country trends and heterogenous levels of entrepreneurship**

**Panel a: Self-employment**

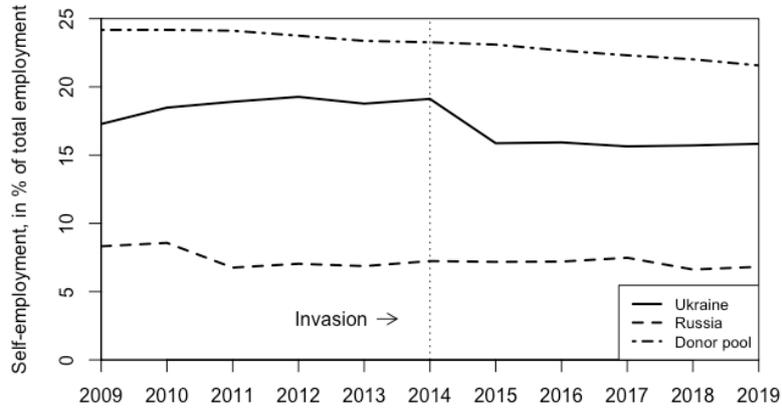

**Panel b: SMEs**

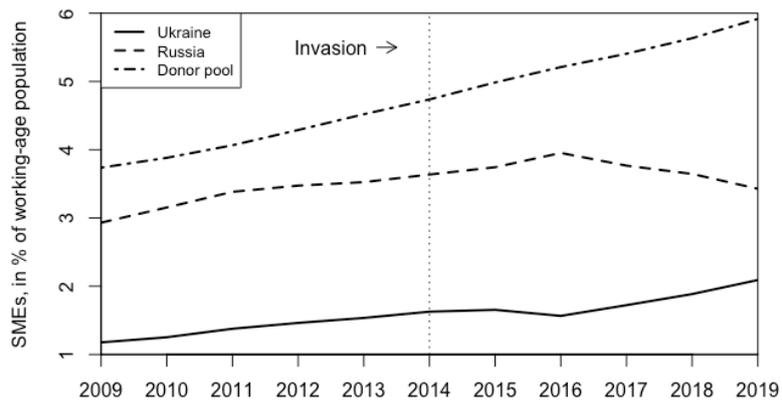

## Table 2: Country statistics and weights in Synthetic Ukraine and Russia

| | % Self-employed of totally employed | | | | % SMEs of working-age population | | | |
|---|---|---|---|---|---|---|---|---|
| | Country mean | Country median | Synthetic Ukraine | Synthetic Russia | Country mean | Country median | Synthetic Ukraine | Synthetic Russia |
| Russia | 7.28 | 7.18 | excluded | - | 3.51 | 3.53 | excluded | - |
| Ukraine | 17.35 | 17.28 | - | excluded | 1.58 | 1.57 | - | excluded |
| Armenia | 42.21 | 42.80 | 0.192 | 0.000 | 2.19 | 2.23 | 0.002 | 0.001 |
| Australia | 17.38 | 17.14 | 0.000 | 0.000 | 13.49 | 13.18 | 0.001 | 0.000 |
| Austria | 13.09 | 13.20 | 0.000 | 0.000 | 1.55 | 1.54 | 0.107 | 0.002 |
| Azerbaijan | 67.69 | 68.06 | 0.000 | 0.000 | 0.83 | 0.78 | 0.048 | 0.001 |
| Belarus | 4.12 | 4.06 | 0.000 | 0.595 | 0.85 | 0.88 | 0.003 | 0.001 |
| Belgium | 14.55 | 14.36 | 0.000 | 0.000 | 7.07 | 7.01 | 0.001 | 0.000 |
| Bulgaria | 12.10 | 12.11 | 0.000 | 0.000 | 2.44 | 2.57 | 0.072 | 0.200 |
| Croatia | 17.53 | 18.08 | 0.000 | 0.000 | 5.64 | 5.60 | 0.027 | 0.000 |
| Cyprus | 16.72 | 17.56 | 0.000 | 0.038 | 29.30 | 28.56 | 0.001 | 0.000 |
| Czech Republic | 17.41 | 17.34 | 0.013 | 0.000 | 5.18 | 5.08 | 0.004 | 0.000 |
| Denmark | 8.96 | 9.10 | 0.000 | 0.000 | 7.39 | 6.94 | 0.038 | 0.363 |
| Estonia | 9.14 | 9.11 | 0.374 | 0.000 | 16.46 | 16.03 | 0.000 | 0.000 |
| Finland | 13.53 | 13.44 | 0.000 | 0.000 | 6.87 | 6.91 | 0.001 | 0.000 |
| France | 11.41 | 11.54 | 0.000 | 0.000 | 4.66 | 4.56 | 0.001 | 0.000 |
| Georgia | 55.97 | 56.53 | 0.000 | 0.000 | 5.73 | 5.40 | 0.002 | 0.000 |
| Germany | 11.05 | 11.20 | 0.000 | 0.000 | 1.04 | 1.07 | 0.000 | 0.001 |
| Hungary | 11.42 | 11.28 | 0.000 | 0.000 | 5.55 | 5.79 | 0.003 | 0.001 |
| Iceland | 12.59 | 12.72 | 0.000 | 0.000 | 14.81 | 14.43 | 0.002 | 0.000 |
| India | 80.34 | 80.12 | 0.000 | 0.000 | 0.13 | 0.13 | 0.154 | 0.167 |
| Ireland | 16.56 | 16.81 | 0.000 | 0.000 | 6.21 | 6.02 | 0.001 | 0.000 |
| Israel | 11.97 | 11.99 | 0.000 | 0.000 | 6.53 | 6.54 | 0.007 | 0.001 |
| Italy | 24.51 | 24.82 | 0.000 | 0.000 | 3.66 | 3.58 | 0.003 | 0.001 |
| Japan | 11.62 | 11.53 | 0.000 | 0.234 | 0.09 | 0.07 | 0.001 | 0.025 |
| Korea, Rep. | 27.47 | 27.41 | 0.000 | 0.000 | 1.39 | 1.33 | 0.004 | 0.001 |
| Lithuania | 11.67 | 11.77 | 0.000 | 0.119 | 5.28 | 5.46 | 0.029 | 0.000 |
| Mexico | 32.71 | 32.92 | 0.000 | 0.000 | 0.77 | 0.67 | 0.000 | 0.005 |
| Mongolia | 52.51 | 50.72 | 0.000 | 0.000 | 4.54 | 4.65 | 0.002 | 0.000 |
| Montenegro | 18.99 | 18.85 | 0.000 | 0.000 | 6.95 | 6.86 | 0.004 | 0.000 |
| North Macedonia | 26.56 | 26.84 | 0.001 | 0.000 | 5.76 | 6.04 | 0.004 | 0.001 |
| Norway | 7.15 | 7.02 | 0.000 | 0.013 | 7.63 | 7.47 | 0.002 | 0.000 |
| Peru | 55.44 | 55.17 | 0.000 | 0.000 | 5.07 | 5.12 | 0.003 | 0.000 |
| Poland | 21.79 | 21.83 | 0.000 | 0.000 | 0.95 | 0.89 | 0.000 | 0.001 |
| Portugal | 20.60 | 21.54 | 0.000 | 0.000 | 8.89 | 8.84 | 0.003 | 0.000 |
| Romania | 30.83 | 32.58 | 0.000 | 0.000 | 7.14 | 6.98 | 0.003 | 0.000 |
| Serbia | 31.22 | 31.35 | 0.000 | 0.000 | 2.43 | 2.33 | 0.003 | 0.216 |
| Singapore | 31.22 | 14.65 | 0.000 | 0.000 | 2.43 | 6.02 | 0.001 | 0.000 |
| Slovak Republic | 15.33 | 15.46 | 0.000 | 0.000 | 5.35 | 5.66 | 0.003 | 0.000 |
| Slovenia | 16.05 | 16.24 | 0.416 | 0.000 | 4.56 | 4.65 | 0.001 | 0.000 |
| Spain | 16.99 | 16.94 | 0.000 | 0.000 | 7.37 | 7.33 | 0.002 | 0.000 |
| Sweden | 10.33 | 10.41 | 0.000 | 0.000 | 7.35 | 7.31 | 0.004 | 0.000 |
| Switzerland | 15.25 | 15.27 | 0.000 | 0.000 | 6.63 | 6.61 | 0.003 | 0.000 |
| Tajikistan | 36.86 | 37.29 | 0.000 | 0.000 | 0.36 | 0.41 | 0.445 | 0.003 |
| United Kingdom | 14.59 | 14.63 | 0.000 | 0.000 | 7.88 | 7.49 | 0.001 | 0.000 |
| Uzbekistan | 40.13 | 40.32 | 0.000 | 0.000 | 0.53 | 0.51 | 0.001 | 0.002 |

**Figure 2: The conflict's effect on self-employment in Ukraine vs. Synthetic Ukraine**

**Panel a: Trends in self-employment for Ukraine vs. Synthetic Ukraine**

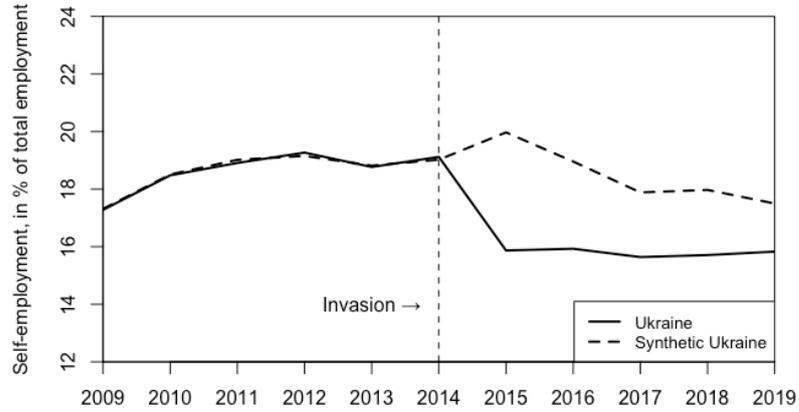

**Panel b: Self-employment gap for Ukraine vs. Synthetic Ukraine**

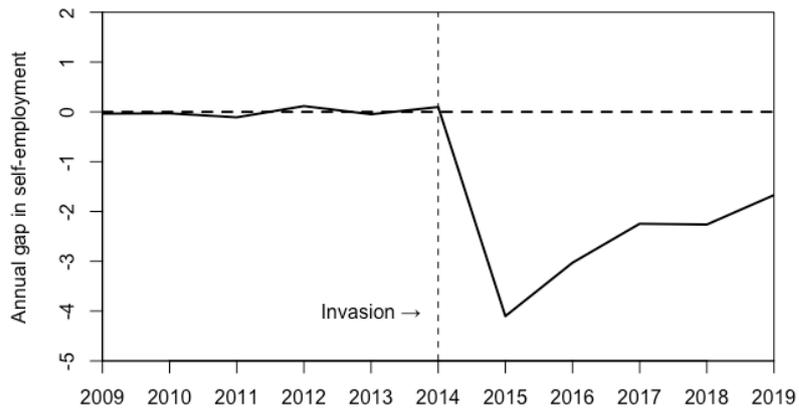

**Panel c: Placebo tests for self-employment gaps in Ukraine and donor-pool countries**

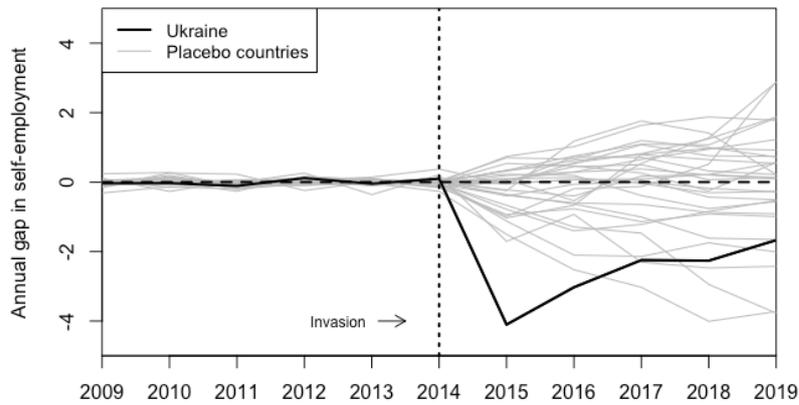

**Figure 3: The conflict's effect on SMEs in Ukraine and Russia vs. their synthetic versions**

**Panel a: Trends in SMEs**

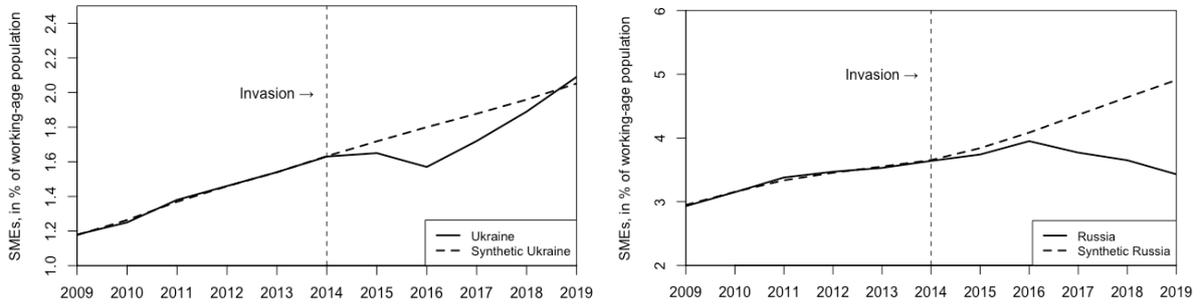

**Panel b: SME gaps**

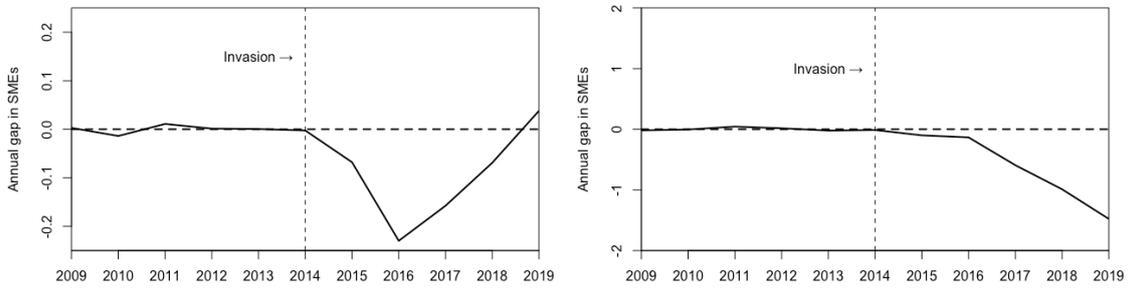

**Panel c: Placebo tests for SME gaps and donor-pool countries**

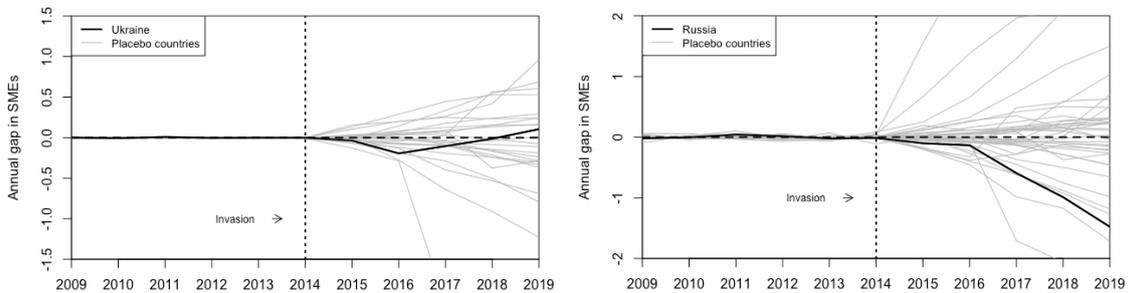

**Figure 4: Birth and death rates of SMEs in Russia vs. Synthetic Russia**

**Panel a: Trends in the SME birth rate**

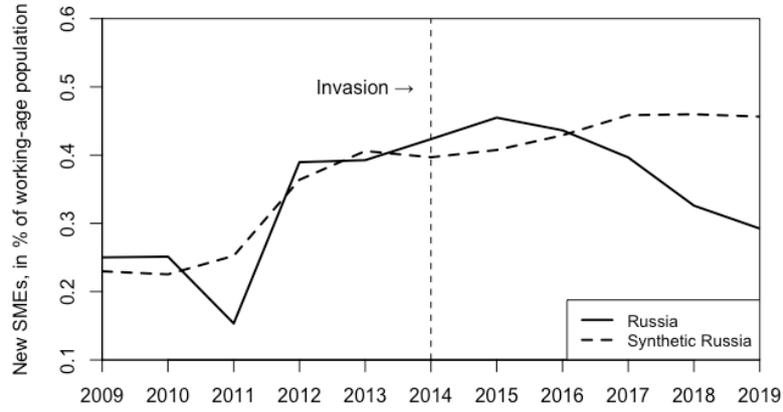

**Panel b: Trends in the SME death rate**

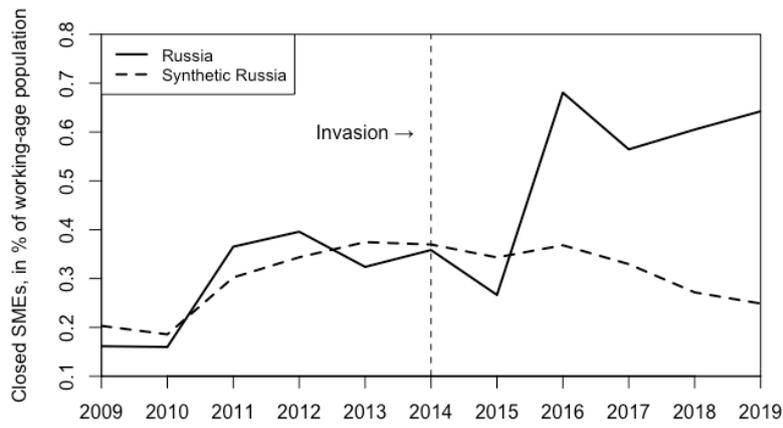

**Appendix**

**Table A.1: Variable definitions, data sources, and summary statistics**

| Variable | Description | Data source | Mean | SD | Q1 | Median | Q3 | Pre-inv. mean | Post-in. mean | Δ in means |
|---|---|---|---|---|---|---|---|---|---|---|
| % Self-employed | Share of total employment, modelled ILO estimate. Self-employed are defined as workers whose remuneration is directly dependent upon profits derived from goods&services produced. | International Labor Organization, ILOSTAT | 22.45 | 16.58 | 12.01 | 15.88 | 26.84 | 23.12 | 21.65 | -1.47 |
| % SMEs | Business density is defined as the total number of registered firms per 1.000 working-age people (those ages 15–64). The units of measurement are private, formal sector companies with limited liability. | World Bank, Entrepreneurship Database | 5.68 | 5.19 | 1.69 | 5.41 | 7.01 | 5.12 | 6.30 | 1.18*** |
| GDP per capita | GDP per capita is gross domestic product divided by mid-year population. Data are in current U.S. dollars. | World Bank, National accounts | 26,509 | 22,699 | 6,988 | 18,656 | 43,647 | 26,336 | 26,717 | 381 |
| GDP growth | Annual percentage growth rate of GDP at market prices based on constant local currency. Aggregates are based on constant 2015 prices, expressed in U.S. dollars | World Bank, National accounts data | 2.30 | 3.84 | 0.99 | 2.43 | 4.10 | 1.75 | 2.96 | 1.20*** |
| Foreign direct investment, net inflows | Foreign direct investment refers to direct investment equity flows in the reporting economy. It is the sum of equity capital, reinvestment of earnings, and other capital. Data in current US$. | World Bank, International Monetary Fund | 17,021 | 35,107 | 1,037 | 4,769 | 21,396 | 16,658 | 17,456 | 798 |
| Starting a business - Score | The score for starting a business is the simple average of the scores for each of the component indicators: the procedures, time and cost for an entrepreneur to start and formally operate a business, as well as the paid-in minimum capital requirement. | World Bank, Doing Business project | 86.85 | 7.91 | 83.10 | 89.00 | 91.90 | 84.92 | 89.17 | 4.25*** |
| Domestic credit to private sector | Domestic credit to private sector, expressed as percentage of GDP. The measure refers to financial resources provided to the private sector by financial corporations, such as through loans, purchases of nonequity securities, and trade credits and other accounts receivable, that establish a claim for repayment. | World Bank, International Monetary Fund | 70.60 | 49.63 | 44.98 | 65.10 | 115.70 | 84.02 | 76.50 | -7.52* |
| Unemployment | Share of total employment, modelled ILO estimate. Unemployment refers to the share of the labor force that is without work but available for and seeking employment. | International Labor Organization, ILOSTAT | 8.79 | 5.51 | 5.10 | 7.01 | 10.70 | 9.38 | 8.10 | -1.28*** |
| Research and development expenditure | Gross domestic expenditures on R&D, expressed as a percent of GDP. They include both capital and current expenditures in the four main sectors: Business enterprise, Government, Higher education and Private non-profit. R&D covers basic research, applied research, and experimental development. | World Bank, World Development Indicators (WDI) | 1.44 | 1.13 | 0.49 | 1.18 | 2.19 | 1.42 | 1.47 | 0.04 |
| Patent applications, residents | Patent applications are worldwide patent applications filed through the Patent Cooperation Treaty procedure or with a national patent office for exclusive rights for an invention. | World Bank, WDI | 12,422 | 44,473 | 136 | 792 | 2,288 | 12,540 | 12,325 | -215 |